\documentclass[aps,prb,twocolumn,superscriptaddress,footinbib,amsmath,amssymb]{revtex4} %%191227 made it prb (instead of a!)
\usepackage{amsmath}
\usepackage{gensymb}
\usepackage{dcolumn}% Align table columns on decimal point
\usepackage{bm}% bold math
\usepackage[pdftex]{graphicx,color} % process figs with pdflatex or pdftex
\usepackage{epstopdf} 
\usepackage{graphicx}
\usepackage{float}
\usepackage[english]{babel}
\usepackage[normalem]{ulem} 

\usepackage{array}
\newcolumntype{P}[1]{>{\centering\arraybackslash}p{#1}}
\newcolumntype{M}[1]{>{\centering\arraybackslash}m{#1}}

\begin{document}
%%\title{Reflectivity_of_Finite_3D_GaAs_Photonic_Band_Gap_Crystals}
\title{Reflectivity of Finite 3D GaAs Photonic Band Gap Crystals}

\author{Takeyoshi Tajiri}
\affiliation{Institute of Industrial Science, The University of Tokyo, 4-6-1 Komaba, Meguro-ku, Tokyo 153-8505, Japan}
\affiliation{Department of Computer and Network Engineering, University of Electro-Communications, 1-5-1 Chofugaoka, Chofu, Tokyo 182-8585, Japan}

\author{Shun Takahashi}
\affiliation{Kyoto Institute of Technology, Matsugasaki, Sakyo-ku, Kyoto 606-8585, Japan}

\author{Cornelis A.M. Harteveld}
\affiliation{Complex Photonic Systems (COPS), MESA+ Institute for Nanotechnology, University of Twente, P.O. Box 217, 7500 AE Enschede, The Netherlands}

\author{Yasuhiko Arakawa}
\affiliation{Institute for Nano Quantum Information Electronics, The University of Tokyo, 4-6-1 Komaba, Meguro-ku, Tokyo 153-8505, Japan}

\author{Satoshi Iwamoto}
\affiliation{Institute of Industrial Science, The University of Tokyo, 4-6-1 Komaba, Meguro-ku, Tokyo 153-8505, Japan}
\affiliation{Research Center for Advanced Science and Technology, Institute of Industrial Science, The University of Tokyo, 4-6-1 Komaba, Meguro-ku, Tokyo 153-0041, Japan}

\author{Willem L. Vos}
\affiliation{Complex Photonic Systems (COPS), MESA+ Institute for Nanotechnology, University of Twente, P.O. Box 217, 7500 AE Enschede, The Netherlands}

\date{January 16th, 2020; in preparation for Physical Review B}

\begin{abstract}
We study the optical reflectivity of three-dimensional (3D) photonic band gap crystals with increasing thickness. 
The crystals consist of GaAs plates with nanorod arrays that are assembled by an advanced stacking method into high-quality 3D woodpile structures. 
We observe intense and broad reflectivity peak with stop bands that correspond to a broad gap in the photonic band structures. 
The maximum reflectivity quickly reaches high values even for a few crystal layers. 
Remarkably, the bandwidth of the stop bands hardly decreases with increasing crystal thickness, in good agreement with FDTD simulations. 
This behavior differs remarkably from the large changes observed earlier in weakly interacting 3D photonic crystals. 
The nearly constant bandwidth and high reflectivity are rationalized by multiple Bragg interference that occurs in strongly interacting photonic band gap crystals, whereby the incident light scatters from multiple reciprocal lattice vectors simultaneously, 
in particular from oblique ones that are parallel to a longer crystal dimension and thus experience hardly any finite size effects. 
Our new insights have favorable consequences for the application of 3D photonic band gap crystals, notably since even thin structures reveal the full band gap functionality, including devices that shield quantum bits from vacuum fluctuations. 
\end{abstract}

\maketitle

%%%%%%%%%%%%%%%%%%%%%%%%%%  body  %%%%%%%%%%%%%%%%%%%%%%%%%%

%%%%%%%%%%%%%%%%%%%%%%%%%%
\section{Introduction}
%%%%%%%%%%%%%%%%%%%%%%%%%%
There is a worldwide interest in three-dimensional (3D) photonic crystals that radically control both the propagation and the emission of light~\cite{Bykov1972SovPhysJETP,Yablonovitch1987PRL,John1987PRL,John1990PRL,Lopez2003AdvMater,Lourtioz2008Book,Joannopoulos2008Book,Ghulinyan2015Book}. 
In photonic crystals the dielectric function varies spatially with a periodicity on length scales comparable to the wavelength of light. 
Due to the long-range periodic order, the photonic dispersion relations are organized in bands, analogous to electron bands in a semiconductor~\cite{Ashcroft1976Book}. 
When the frequency of light lies in a gap in the dispersion relations for a certain wave vector tending from the origin to the Brillouin zone boundary, light cannot propagate in the corresponding direction as a result of Bragg diffraction~\cite{Ashcroft1976Book}. 
Such a directional gap or stop gap is usually probed with reflection or transmission experiments where a reflectivity peak or transmission trough occurs, also known as a stop band~\cite{Lin1998Nature, Thijssen1999PRL, Noda1999APL, Noda2000Science, Blanco2000Nature, Vlasov2000Nature, Palacios-Lidon2002APL, Aoki2003NatMater, Schilling2005APL, Garcia2007AM, Subramania2007OE, Takahashi2009NatMat, Staude2010OL, Huisman2011PRB, Subramania2011NL, Frolich2013AM, Marichy2016SR, Adhikary2019Arxiv}. 
To first order, the width of a stop band is proportional to the ratio of the dominant Fourier component of the dielectric function and the average dielectric constant~\cite{Yariv1980book,Vos2015book}. 
The width thus serves as a gauge of a generalized photonic interaction strength between the light and the photonic crystal nanostructure~\cite{Vos1996JPCM} that notably depends on the dielectric contrast, the volume fraction of high-dielectric material, and a relevant structure factor~\cite{Vos2015book}.\footnote{The photonic interaction strength $S$ is formally defined as the polarizability $\alpha$ per volume $V$ of each building block (or scatterer): $S \equiv \alpha / V$~\cite{Vos1996PRB, Vos2015book}.} 
When the photonic interaction strength exceeds a certain threshold, a 3D photonic band gap can emerge, a frequency range for which light modes are forbidden for \textit{all} wave vectors and \textit{all} polarizations~\cite{Bykov1972SovPhysJETP,Yablonovitch1987PRL,John1987PRL,John1990PRL,Lopez2003AdvMater,Lourtioz2008Book,Joannopoulos2008Book,Ghulinyan2015Book}. 
Since the local density of states also vanishes, the photonic band gap is a powerful tool to radically control spontaneous emission and cavity quantum electrodynamics (QED) of embedded quantum emitters~\cite{Bykov1972SovPhysJETP, John1990PRL,  Lambropoulos2000RP, Vos2015book}. 
Applications of 3D photonic band gap crystals range from dielectric reflectors for antennae~\cite{Smith1998MOTL} and for efficient photovoltaic cells~\cite{Bermel2007OE, Wehrspohn2012JO, Koenderink2015science}, via white light-emitting diodes~\cite{Nelson2011NM, David2012RPP}, to elaborate 3D waveguides~\cite{Li2003JOSA} for 3D photonic integrated circuits~\cite{Staude2011OptLett, Ishizaki2013NP, Tajiri2019Optica}, and to low-threshold miniature lasers~\cite{Tandaechanurat2011NP} and devices to control quantum noise for quantum measurement, amplification, and information processing~\cite{Clerk2010RMP, Leistikow2011PRL, Vos2015book}.

Although considerable progress has been made in experimental studies on real 3D photonic crystals, physical understanding of photonic band gaps is mostly based on theories that pertain to infinite and perfect crystals~\cite{Joannopoulos2008Book,Lourtioz2008Book}. 
This theoretical situation is obviously different from real photonic crystals and photonic crystal devices that are finite in extent and surrounded by free space, also known as a crystal with finite support~\cite{wikipedia2018support}. 
Remarkably, the understanding of photonic crystals with finite support is much less developed. 
Regarding densities of states, it has been reported for 3D photonic band gap crystals that the LDOS in the gap at the center of the crystal seems to decrease exponentially with crystal size~\cite{Hermann2002JOSAB, Kole2003thesis}. 
The scaling of the volume averaged LDOS (the DOS) has been studied by Hasan \textit{et al.} who found it to decrease linearly with increasing crystal size $L$~\cite{Hasan2018PRL}. 

Finite crystal size is also known to affect the propagation of light. 
In previous work, the stop bandwidths of weakly interacting photonic crystals (in absence of band gaps) was found to decrease with increasing the crystal thickness~\cite{Bertone1999PRL, Hartsuiker2008thesis}.
Remarkably, however, such finite-size effect for stop band widths has hardly been discussed in strongly interacting photonic crystals with photonic band gaps.
%{\color{red}Studies on finite thickness effect of stop bands in strongly interacting photonic crystals could contribute cultivating novel practical insight of stop bands which have been more eagerly studied in various thin photonic nanostructures such as high contrast gratings.
%At the current situation for studies on stop bands of strongly interacting photonic crsytals, it is inferred that the bands hardly shift with angle of incidence (or incident wave vector) by a numerical work~\cite{Devashish2017PRB}.
%However, the dependence of the widths of stop bands on the size of 3D photonic crystals has to the best of our knowledge not been investigated.}
From numerical work, it is inferred that stop bands hardly shift with angle of incidence (or incident wave vector)~\cite{Devashish2017PRB}, but the dependence of the widths of stop bands on the size of 3D photonic crystals has to the best of our knowledge not been investigated. 
Therefore, we present here a combined experimental, numerical and theoretical study of the stop bands of 3D photonic band gap crystals with different thicknesses.

%%%%%%%%%%%%%%%%%%%%%%%%%%
\section{Methods}
%%%%%%%%%%%%%%%%%%%%%%%%%%
%%%%%%%%%%%%%%%%%%%%%%%%%%
\subsection{3D GaAs woodpile crystals}
%%%%%%%%%%%%%%%%%%%%%%%%%%

%%%%%%%%%%%%%%%%%%%%%%%%%%
\begin{figure}[h!]
\centering
\includegraphics[width=1.0\columnwidth]{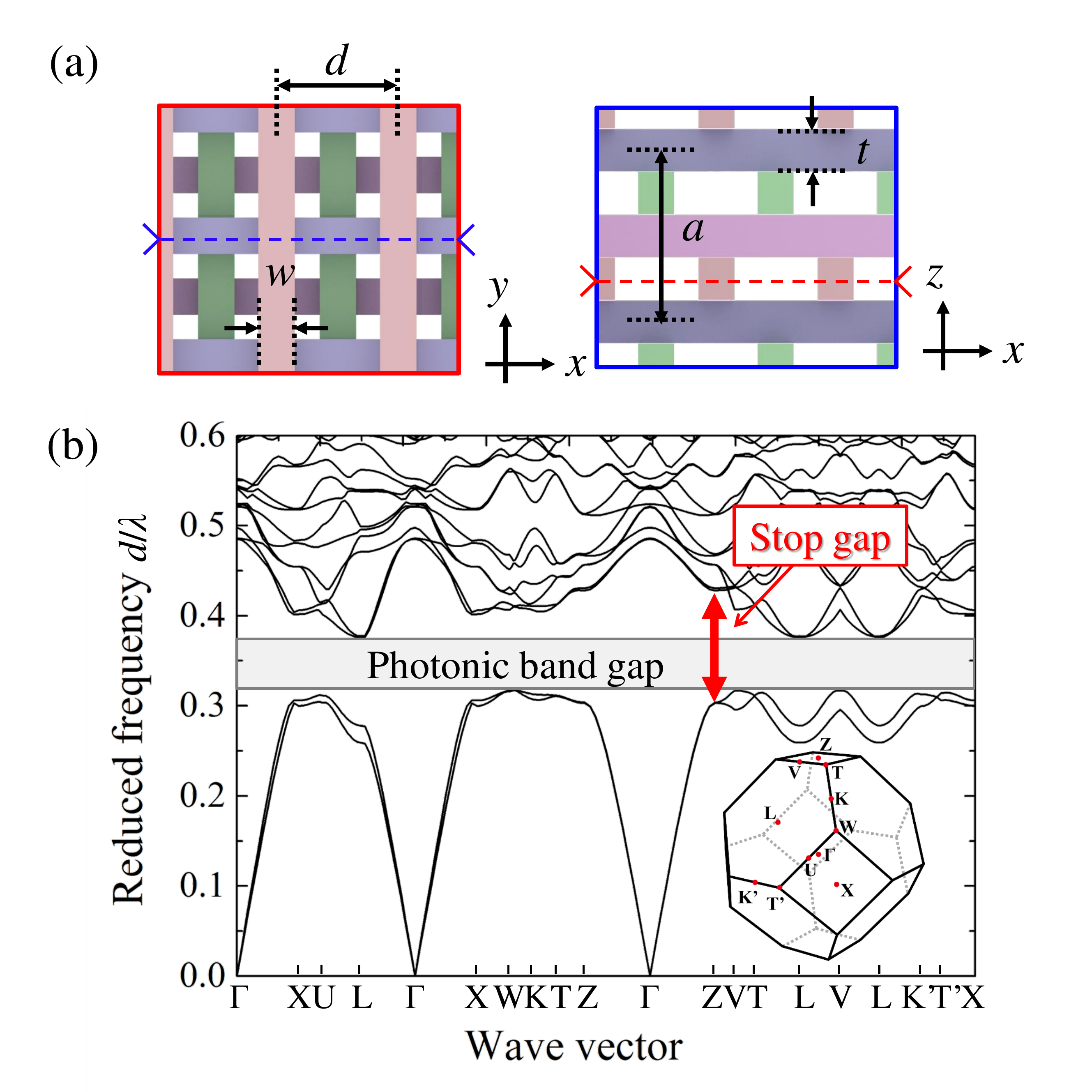}
\caption{(a) Schematic of the woodpile 3D photonic band gap crystal structure. 
Left: view down the $z$-axis. 
Right: $x-z$ cross-section perpendicular to the rod layers at the dashed line in the left image.
The ($x,y$)-lattice parameter $d$, rod width $w$, thickness $t$, and $z$-lattice parameter $a$ are shown. 
The four different colors serve to indicate the four different arrays that are stacked consecutively in the $z$-direction. 
The dashed lines indicate the position of the cross-sectional planes in the neighboring schematics. 
(b) Photonic band structures for our 3D woodpile photonic band gap crystal with frequency reduced by the in-plane lattice parameter ($d/\lambda$).
The wave vector runs between high symmetry points 
where the range has been expanded to include in-plane wave vectors.
The arrow shows the width of the $\Gamma Z$ stop gap.
The inset shows the first Brillouin zone and the relevant high-symmetry points.
}
\label{fig:schematic_structure}
\end{figure} 
%%%%%%%%%%%%%%%%%%%%%%%%%%

Figure~\ref{fig:schematic_structure}(a) shows a schematic of the structure of the 3D photonic crystals studied here. 
The crystals have the well-known woodpile structure where four different arrays of nanorods are stacked along the $z$-direction~\cite{Ho1994SSC}, as shown in the right panel of  Fig.~\ref{fig:schematic_structure}(a). 
In each array, the rectangular nanorods with width $w$ and thickness $t$ are arranged with ($x,y$)-lattice parameter $d$ (left panel). 
In the $z$-direction, the nanorods in neighboring layers are perpendicular to each other. 
nanorods in second-nearest neighbor layers are parallel and shifted by half a period $d/2$. 

Compared to the conventional cubic (non-primitive) unit cell of the diamond structure~\cite{Ashcroft1976Book}, the $(x,y,z)$ coordinate system of our 3D photonic crystal, shown in Fig.~\ref{fig:schematic_structure}, has a face-centered-tetragonal (fct) lattice with $x$-axis unit vector $a_{1} = \frac{1}{\sqrt{2}}[d~d~0]$, $y$-axis $a_{2} = \frac{1}{\sqrt{2}}[\bar{d}~d~0]$, and $z$-axis $a_{3} = [0~0~a]$. 
Since the lattice parameter $d$ in the ($x,y$)-directions turns out to be equal to $d = d_{hkl=110} = D \times d_{001} = D \times a/\sqrt{2}$, with $D \neq 1$, the crystals have a small tetragonal distortion from cubic symmetry (where $D = 1$) that hardly affects the band gap. 

The crystals are made of GaAs and have $d = 470$ nm, $w = 178$ nm and $t = 150$ nm hence $D = 1.11$, resulting in gaps in the near infrared and telecom ranges. 
Figure ~\ref{fig:schematic_structure}(b) shows the photonic band structure calculated for the woodpile crystal structure using the plane-wave expansion method~\cite{Leung1993book, Joannopoulos2008Book}. 
The number of plane waves used in this calculation is 262144, namely $32\times32\times32$ reciprocal lattice vectors in all 3 dimensions, times $2^3$ for the two polarizations at each lattice vector. 
The structure was taken to have the parameters $(w,t,d,a)$ from the design, and the dielectric constant was taken as $\epsilon = 11.56$, typical for GaAs in the telecom range. 
The 3D photonic band gap shown in Fig.~\ref{fig:schematic_structure}(b) ranges from reduced frequency $d/\lambda = 0.318$ to $0.376$, which corresponds to wave numbers between $\nu/c = 6770$ cm$^{-1}$ and $8000$ cm$^{-1}$, corresponding to wavelengths between $1480$~nm and $1250$~nm, which overlaps with the telecom O-band. 
The stop bandwidth in the $\Gamma Z$ direction, that is probed in the experiment, appears between $d/\lambda = 0.303$ and $0.428$, which corresponds to $\nu/c = 6450$ cm$^{-1}$ to $9110$ cm$^{-1}$. 

%%%%%%%%%%%%%%%%%%%%%%%%%%
\begin{figure}[h!]
\centering
\includegraphics[width=1.0\columnwidth]{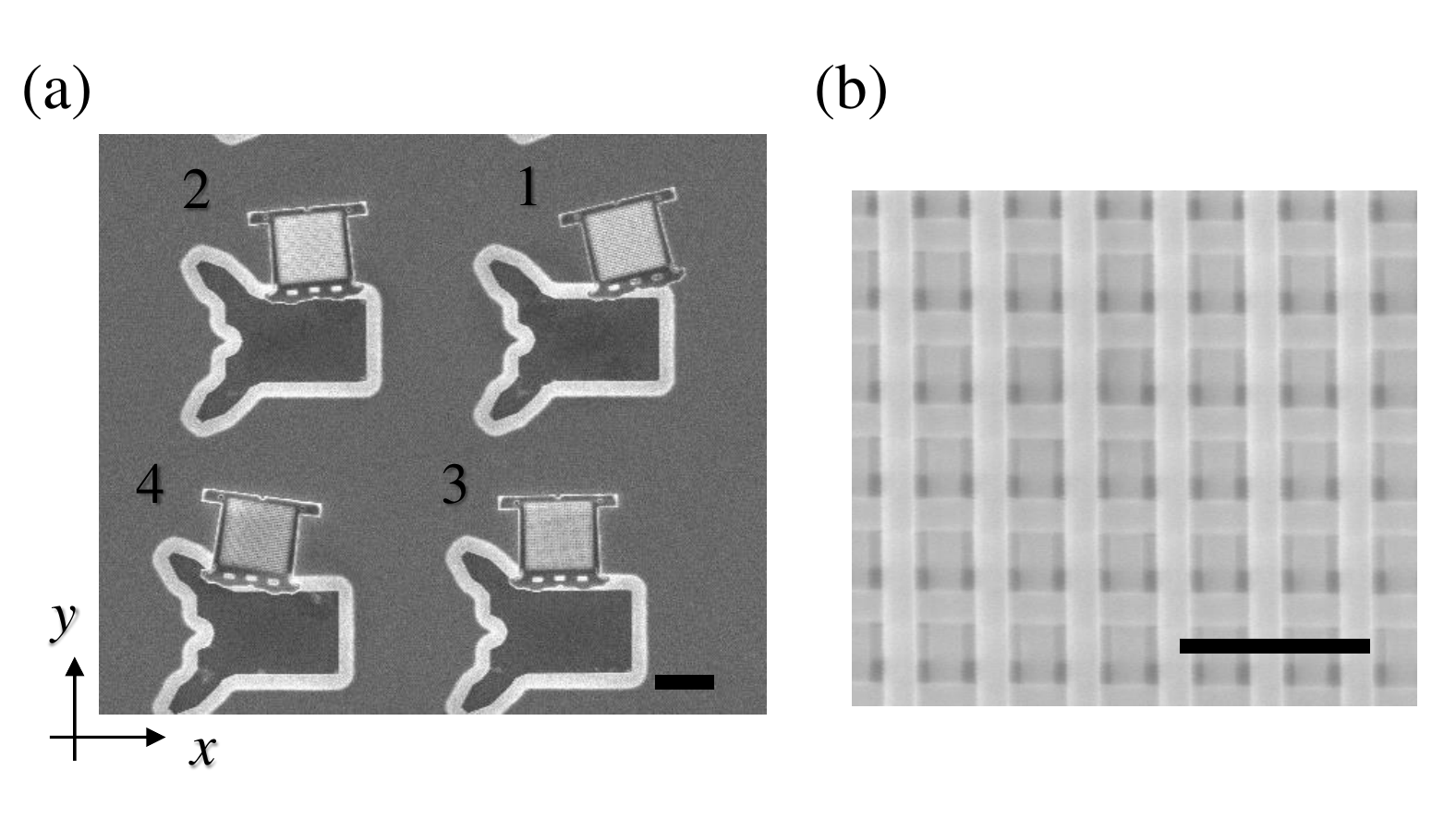}
\caption{SEM images of several GaAs 3D photonic band gap crystals. 
(a) Four woodpile structures are shown with increasing layer thicknesses $L = 1 d_{002}$, $2 d_{002}$, $3 d_{002}$, and $4 d_{002}$ that reside next to the trenches used to stack the crystal layers.
The scale bar is $10~\mu$m long. 
(b) Magnified view of the woodpile structure with a thickness $L = 2 d_{002}$ (lower left in panel (a)). 
The scale bar is $1~\mu$m long. 
The rod have widths $w = 168 \pm 9$~nm, thickness $t = 151 \pm 5$~nm, and ($x,y$)-lattice parameter $d = 480 \pm 10$~nm where the error bars are estimated from multiple SEM measurements. 
} %%end \caption
\label{fig:SEM_of_the_crystals}
\end{figure}
%%%%%%%%%%%%%%%%%%%%%%%%%%

We have fabricated the 3D photonic crystal structures using a micro-manipulation method~\cite{Aoki2008NP, Tajiri2015APL, Iwamoto2016Phot}. 
The fabricated crystals are shown in Fig.~\ref{fig:SEM_of_the_crystals}. 
The 2D arrays of nanorods are fabricated in plates within an area of 11.3 $~\mu$m by 11.3 $~\mu$m by dry etching following e-beam lithography.\footnote{Effectively, such a single layer of nanorods looks like a high-refractive index contrast grating\cite{ChangHasnain2012AOE}.} 
The plates with the rod arrays are stacked one-by-one in the trenches that are also shown in Fig.~\ref{fig:SEM_of_the_crystals}(a), under scanning-electron-microscope (SEM) observation~\cite{Tajiri2019Optica}. 
After stacking a desired number of plates to form a 3D crystal, the whole crystal is removed from a trench and carefully laid next to it, as shown in Fig.~\ref{fig:SEM_of_the_crystals}(a). 
The substrate below the crystals consists of several layers: first a 225 nm-thick-GaAs top layer, next a $3~\mu$m thick AlGaAs sacrificial layer, and finally a bulk GaAs substrate.%%endblue

A magnified SEM image of stacked rod arrays is shown in Fig.~\ref{fig:SEM_of_the_crystals}(b). 
In this particular structure, four layers are successfully stacked in the woodpile structure. 
The 3D crystals in Fig.~\ref{fig:SEM_of_the_crystals}(a) have increasing crystal thicknesses $L = d_{002}$, $2 d_{002}$, $3 d_{002}$, and $4 d_{002}$, with $d_{002} = a/2$ is the spacing of the $hkl=002$ lattice planes in the $z$-direction.

%%%%%%%%%%%%%%%%%%%%%%%%%%
\subsection{Optical setup}
%%%%%%%%%%%%%%%%%%%%%%%%%%
Optical reflectivity was measured using a home-built microscope setup that employs reflective optics and operates in the near infrared range at wavelengths beyond $800$~nm, see Ref.~\cite{Ctistis2010PRB,Huisman2011PRB}. 
Main components are a supercontinuum white light source (Fianium), a Fourier-transform interferometer (Biorad FTS-6000) that operates with $8$ cm$^{-1}$ spectral resolution, and a reflecting objective ($NA=0.65$) to focus the beam to a few microns inside the photonic crystal domains over the large required range of frequencies. 
Signals were collected in the near infrared spectral range between about $4000$~cm$^{-1}$ and more than $12500$~cm$^{-1}$, corresponding to wavelengths between $2500$ nm and $800$~nm. 
Reflectivity was calibrated by taking the ratio of a sample spectrum with the spectrum measured on a clean gold mirror. 
In the spectra, a narrow range near $9300$ cm$^{-1}$ is omitted since it is disturbed by the pump laser of the supercontinuum source. 
Following Ref.~\cite{Adhikary2019Arxiv}, we estimate the error margins in the stopband edges and therefore the stopband width. 
We estimate the standard deviation in the minimum reflectivity at frequencies below the stopband, and the standard deviation of the maximum reflectivity, that both propagate into the determination  of the gap edges.
%%endblue

%%%%%%%%%%%%%%%%%%%%%%%%%%
\subsection{Numerical modeling}
%%%%%%%%%%%%%%%%%%%%%%%%%%
To compute the optical properties of the finite crystals, we employ the finite-difference time domain (FDTD) method using commercially available code (Synopsis Rsoft FULLWAVE). 
In the simulations, the structure has 1 period size in both x and y directions, and periodic boundary conditions are imposed in both the $x$ and $y$ directions, and in the stacking direction $z$ perfectly matched layers are inserted at 10 $\mu$m far from the surface of the substrate. 
The grid size was set as $dx \times dy \times dz = d/16 \times d/16 \times d_{002}/8$. 
The incident light consists of plane waves propagating along the $z$ direction with a range of frequencies parallel to the sample normal, with the same polarization as in the experiments. 
The incident light is pulsed with a single-cycle pulse duration in order to cover a broad frequency range, with a central frequency $8500$ cm$^{-1}$. 
At this time, it is not feasible to take into account the finite numerical aperture, since adding coherently many fields with many different incident wave vectors (as is the case in a focus) is computationally prohibitively expensive.
The time-dependent electric fields of the reflected light are recorded, and we analyze the spectral response using a Fourier transform. 
The reflectivity is calculated as the ratio of the computed reflected intensity and the incident intensity. 
In our simulations, we set the structural parameters of our 3D photonic crystal to be $d = 470$ nm, $w = 178$ nm and $t = 150$ nm, according to the SEM images of our samples to get the best match with the experiments. 
We also included the substrate in the simulations.
%%{\color{blue}XXX GaAs without dispersion which $\epsilon$ or $n$?? In case of GaAs dispersion (which reference?) XXX}
The refractive index for GaAs was taken to be $n = 3.4$ in absence of dispersion. 
In presence of GaAs dispersion, the refractive index varied from $n = 3.34$ at $5000$ cm$^{-1}$ to $n = 3.65$ at $12000$ cm$^{-1}$, and the extinction coefficient monotonically increased from $0$ at $11700$ cm$^{-1}$ to $0.0662$ at $12000$ cm$^{-1}$~\cite{Skauli2003JAP}. 
The simulations were done with a standard personal computer and took about 24 hours per spectrum. 
%%endblue

To model the effect of disorder, we multiplied the reflectivity output by the FDTD code with an exponential extinction factor exp$(-\ell/L)$ for each wavelength. 
For the extinction length $\ell$ we employ the analytical model by Koenderink \textit{et al.} for a 3D photonic crystal~\cite{Koenderink2005PRB}. 
The underlying root-mean-square structural variation is taken to be $\Delta r = 5$~nm, which is a reasonable estimate of typical roughness and size variations in semiconductor nanophotonic structures. 

%%%%%%%%%%%%%%%%%%%%%%%%%%
\section{Results}\label{sec:results}
%%%%%%%%%%%%%%%%%%%%%%%%%%
\subsection{Experiment}
%%%%%%%%%%%%%%%%%%%%%%%%%%
\begin{figure}[h!]
\centering
\includegraphics[width=1.0\columnwidth]{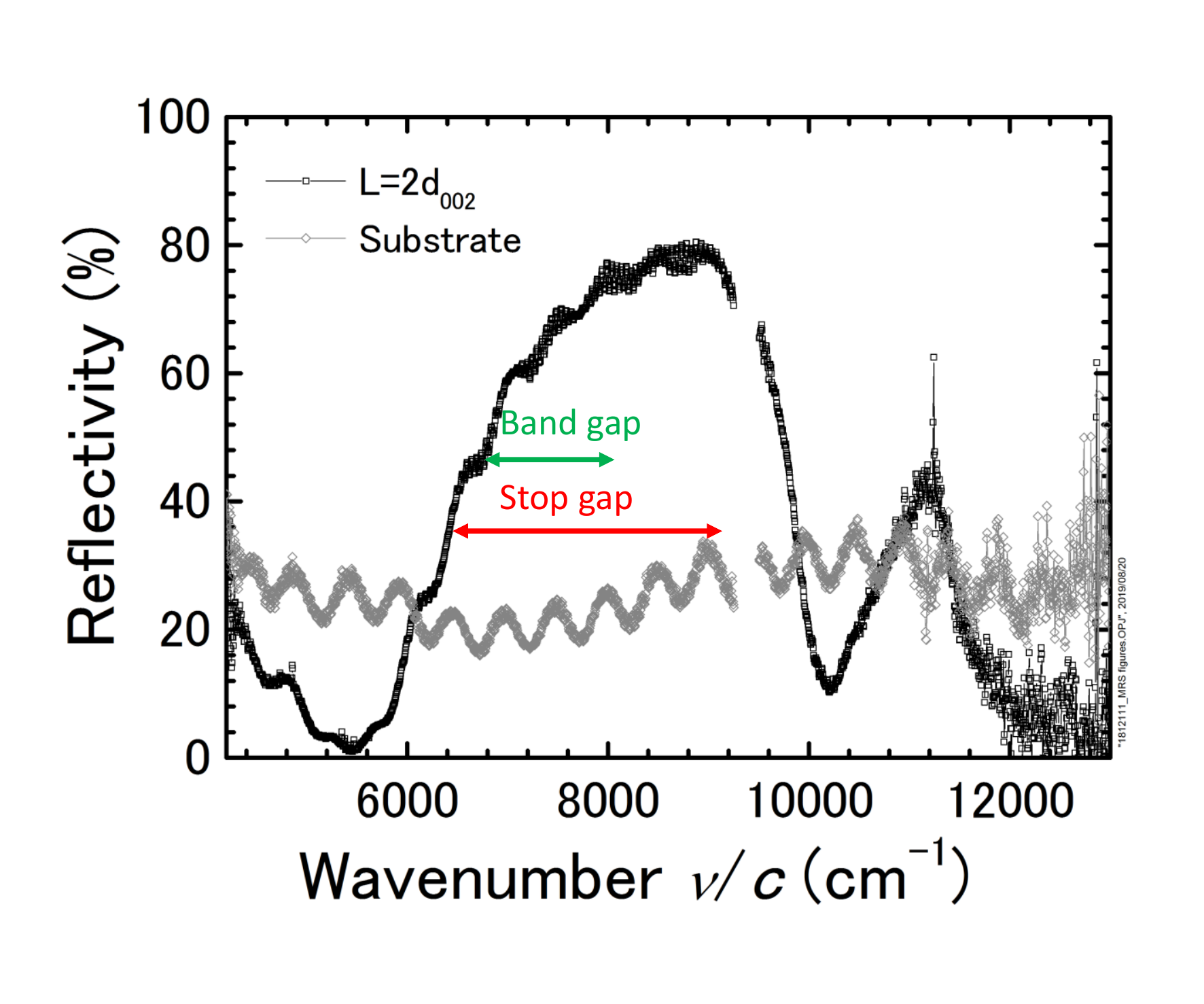}
\caption{Reflectivity versus wavenumber for a 3D photonic crystal with a thickness $L=2d_{002}$ (black circles) and for the substrate (grey circles). 
The range near $9200$ cm$^{-1}$ is excluded as it is disturbed by the pump laser of the supercontinuum white light source. 
The green horizontal arrow indicates the width of the 3D photonic band gap and the red arrow the width of the $\Gamma Z$ stop gap.%%endblue
} 
\label{fig:reflectivity_spectra}
\end{figure}
%%%%%%%%%%%%%%%%%%%%%%%%%%

Figure~\ref{fig:reflectivity_spectra} shows a measured reflectivity spectrum for the 3D photonic crystal of thickness $L=2d_{002}$ together with the spectrum for the substrate.
The reflectivity spectrum of the substrate varies between $20~\%$ and $30~\%$ with clear fringes with a period of about $500$~cm$^{-1}$, and a second, much longer, fringe spacing of about $6000$~cm$^{-1}$.
The fringes are caused by two interference phenomena: 
The $500$~cm$^{-1}$ fringes are caused by interference between the top GaAs layer and the GaAs substrate that are spaced by the $3~\mu$m thick AlGaAs sacrificial layer. 
%%, corresponding to a fringe spacing of about $480$~cm$^{-1}$~\cite{Demtroder2006book}, in good agreement with the observed spacing.
Secondly, the $6000$~cm$^{-1}$ fringes are due to interference inside the $225$~nm thin top layer. 
%%, corresponding to a fringe spacing of $6350$~cm$^{-1}$, in good agreement with the long fringe spacing. 
The 3D photonic crystal reveals an intense and broad reflectivity stop band at frequencies between $6000$~cm$^{-1}$ and $10000$~cm$^{-1}$ that corresponds to the $\Gamma Z$ stop gap in the band structures (\textit{cf.} Fig.~\ref{fig:schematic_structure}). 
The full width at half maximum is about $3300$~cm$^{-1}$, corresponding to a relative bandwidth $\Delta \omega / \omega_c = 41 \%$, which is typical of a strongly interacting photonic crystal~\cite{Vos2015book}, and fairly comparable to results obtained for a silicon woodpile structure~\cite{Euser2008PRB}, which makes sense in view of the similar refractive index contrasts of GaAs-air and of Si-air structures.%%endblue

%%%%%%%%%%%%%%%%%%%%%%%%%%
\begin{figure}[h!]
\centering
\includegraphics[width=1.0\columnwidth]{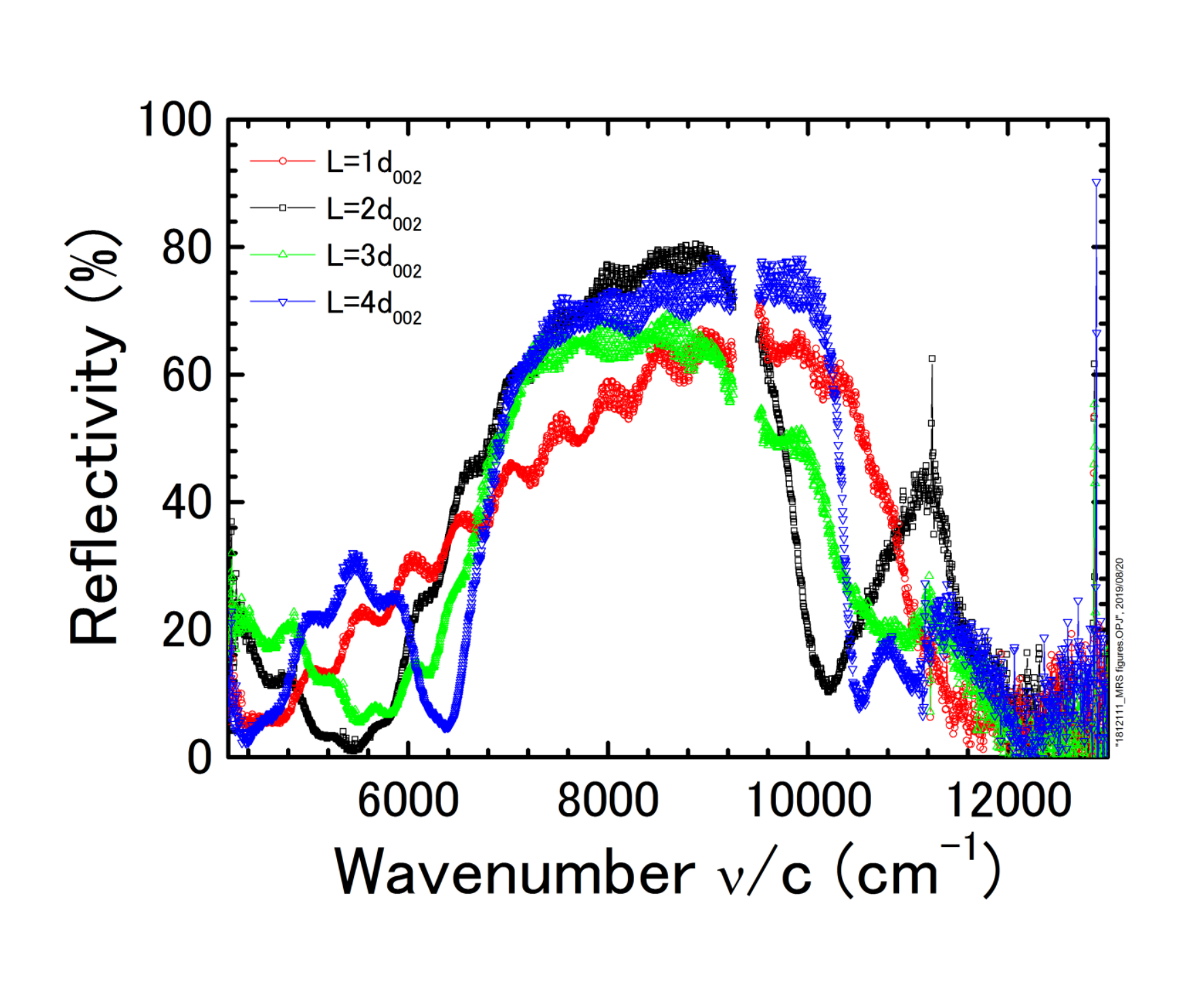}
\caption{Reflectivity versus wavenumber for the crystals with $L = d_{002}$ (red circles), $L = 2 d_{002}$ (black squares), $L = 3 d_{002}$ (green triangles), and $L = 4 d_{002}$ (blue inverted triangles). 
The range near $9200$ cm$^{-1}$ is excluded as it is disturbed by the pump laser of the supercontinuum white light source. 
} 
\label{fig:reflectivity_all_spectra}
\end{figure}
%%%%%%%%%%%%%%%%%%%%%%%%%%

Figure~\ref{fig:reflectivity_all_spectra} shows reflectivity spectra measured for the 3D photonic crystals with increasing thickness $L = d_{002}, 2 d_{002}, 3 d_{002}$ and $4 d_{002}$. 
It is gratifying that all four spectra show a stop band between about $6800$~cm$^{-1}$ and $10000$~cm$^{-1}$ as taken at half height. 
%%The stop band is relatively broad as expected for a strongly interacting photonic crystal. 
There are a number of features that are characteristic of a high crystal quality. 
Firstly, in the stop band the maximum reflectivity is overall quite elevated up to $70$ or even $80 \%$. 
Secondly, since all spectra reveal the $500$~cm$^{-1}$ fringes of the sacrifical layer interference, the transmission of light through the crystals is hardly attenuated by scattering from random disorder. 
Thirdly, all spectra reveal a minimum reflectivity near $0 \%$ below the stop band, which is part of the Fabry-P{\'e}rot interference due to interference between crystal's front and back surface reflections (see, \textit{e.g.}, Ref.~\cite{Devashish2017PRB}), whereas such interference would disappear if the interference would be perturbed by random scattering or by non-planar back surfaces. 

When we zoom in on the spectra in Figure~\ref{fig:reflectivity_all_spectra}, we note several features and differences. 
For the thinnest crystal ($L = d_{002}$), the stop band is both the broadest (from $6400$~cm$^{-1}$ to $10800$~cm$^{-1}$) and the reflectivity the lowest, with a maximum of about $60 \%$ at the stop band center. 
A broadened and less intense spectrum is reasonably expected for a thin crystal. 
%%end blue
In comparison to the spectrum for $L = d_{002}$, higher reflectivity within the stop bands were observed for the other 3D photonic crystals with larger thicknesses. 
In addition, the interference fringes which are clearly observed for $L = d_{002}$ within the range of stop band becomes obscure in other thicker crystals. 
This result indicates that, for the case of large thicknesses, incident light is mostly reflected by the 3D photonic crystals without reaching the substrate underneath, suggesting strong interactions with photons in those 3D photonic crystals. 
In addition to the above-mentioned spectral features which indicate strong interactions with photons, we can also observe several other features originating in unintentional reasons. 
One is that a peak around $5000$~cm$^{-1}$ is observed only for the crystal of $L = 4d_{002}$.
The reason of this peak would be a stacking order in this 3D crystal (see Appendix) that inadvertently differs from the other crystals.
Another one is that at higher wave numbers in the stop band around $9000$~cm$^{-1}$, the reflectivity of the $L = 3d_{002}$ crystal becomes lower than for other thicknesses. 
This is tentatively attributed to inadvertent errors in our 3D crystals, such as misalignment between the stacked layers. 
%%}%%end-red

%%%%%%%%%%%%%%%%%%%%%%%%%%
\subsection{Numerical simulations}
%%%%%%%%%%%%%%%%%%%%%%%%%%
\begin{figure}[h!]
\centering
\includegraphics[width=1.0\columnwidth]{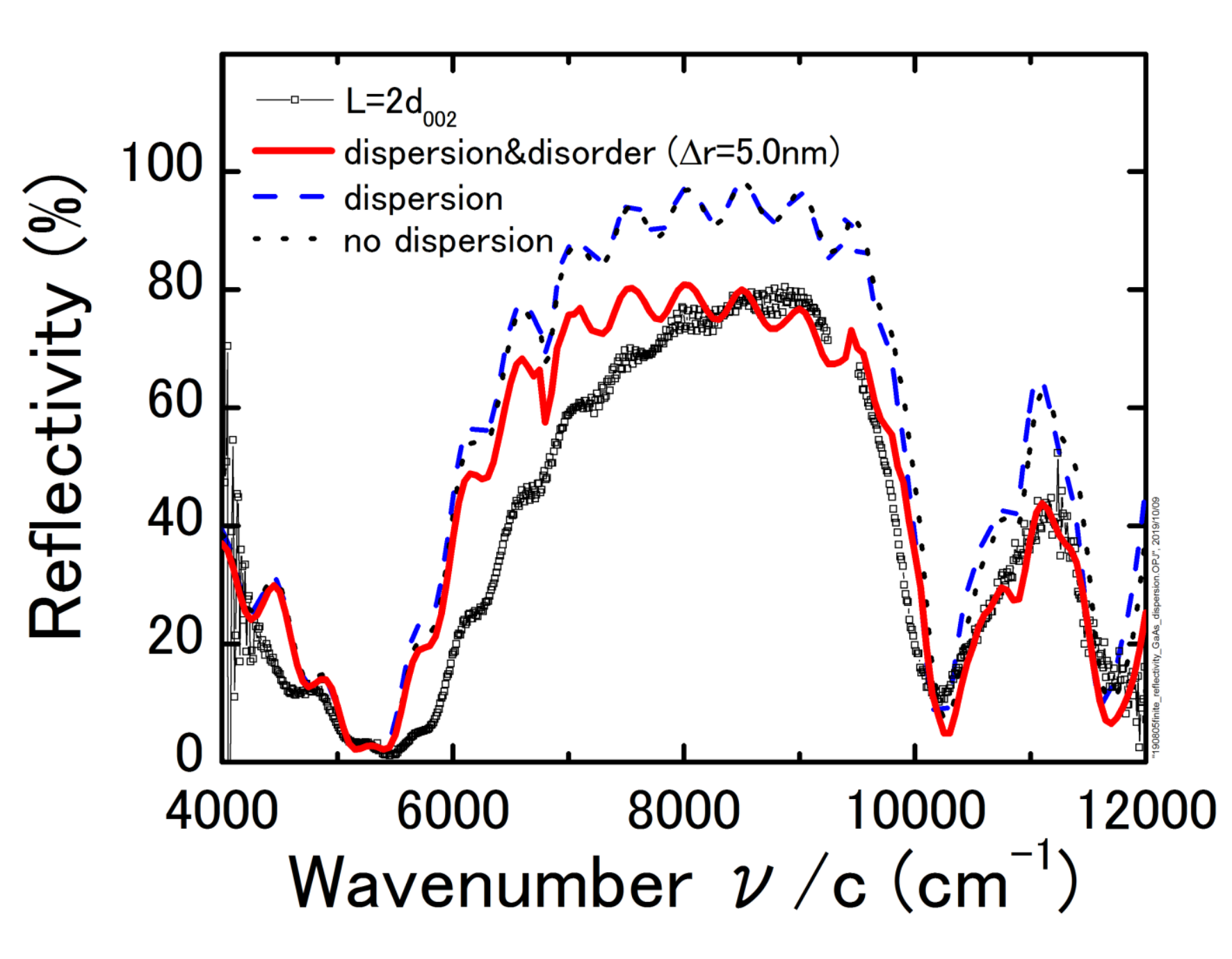}
\caption{Reflectivity versus wavenumber calculated for a thickness $L = 2d_{002}$ by 3D finite-difference time domain simulations for the E-field parallel to the $X$-axis. 
The blue dashed curve takes into account GaAs dispersion, the black dotted curve is in absence of dispersion, and the red drawn curve takes into account both dispersion~\cite{Skauli2003JAP} and disorder~\cite{Koenderink2005PRB}.
Experimental data for $L = 2d_{002}$ are shown as black connected squares for comparison.} 
\label{fig:FDTD_versus_experiment}
\end{figure}
%%%%%%%%%%%%%%%%%%%%%%%%%%

To interpret our experiments, we have computed reflectivity spectra using FDTD simulations. 
In Figure~\ref{fig:FDTD_versus_experiment}, calculated reflectivity spectra are compared with the measured one for a crystal thickness $L = 2d_{002}$. 
For a first set of simulations where the GaAs refractive index is constant and non dispersive (black dotted curve in Fig.~\ref{fig:FDTD_versus_experiment}) the width of the reflectivity peak matches well with the measured one, which indicates that the peak originates in a stop band of our 3D photonic crystal. 
%which confirms that the structural parameters are a faithful representation of the crystal.
The fringes due to the interference in our substrate also match well in period with the measured fringes. 
On the other hand, it is apparent that the simulated spectrum has a substantially higher reflectivity (up to $99 \%$) than the measured one. 
To interpret this marked difference in reflectivity, we explored the roles of dispersion of the GaAs refractive index and of structural disorder. %%endblue

When we incorporate dispersion in the FDTD simulations, the reflectivity spectrum hardly changes as shown in Fig.~\ref{fig:FDTD_versus_experiment} (blue curve). 
Therefore, we reject the hypothesis that dispersion is important. 
In contrast, when the disorder was taken into account, the reflectivity is reduced around the higher frequency side of the stop band, which results in close matching between experiment and simulation at the higher frequency range.
This result suggests that the disorder in our 3D photonic crystals is a dominant reason for the reduction in the reflectivity at the high frequency side of the stop band.
Although the lower frequency side of the stop band is still mismatched, this may be caused by other factors excluded in our simulation such as the effect of finite numerical aperture of our reflection objective and the misalignment between stacked layers.

%%%%%%%%%%%%%%%%%%%%%%%%%%
\begin{figure}[h!]
\centering
\includegraphics[width=1.0\columnwidth]{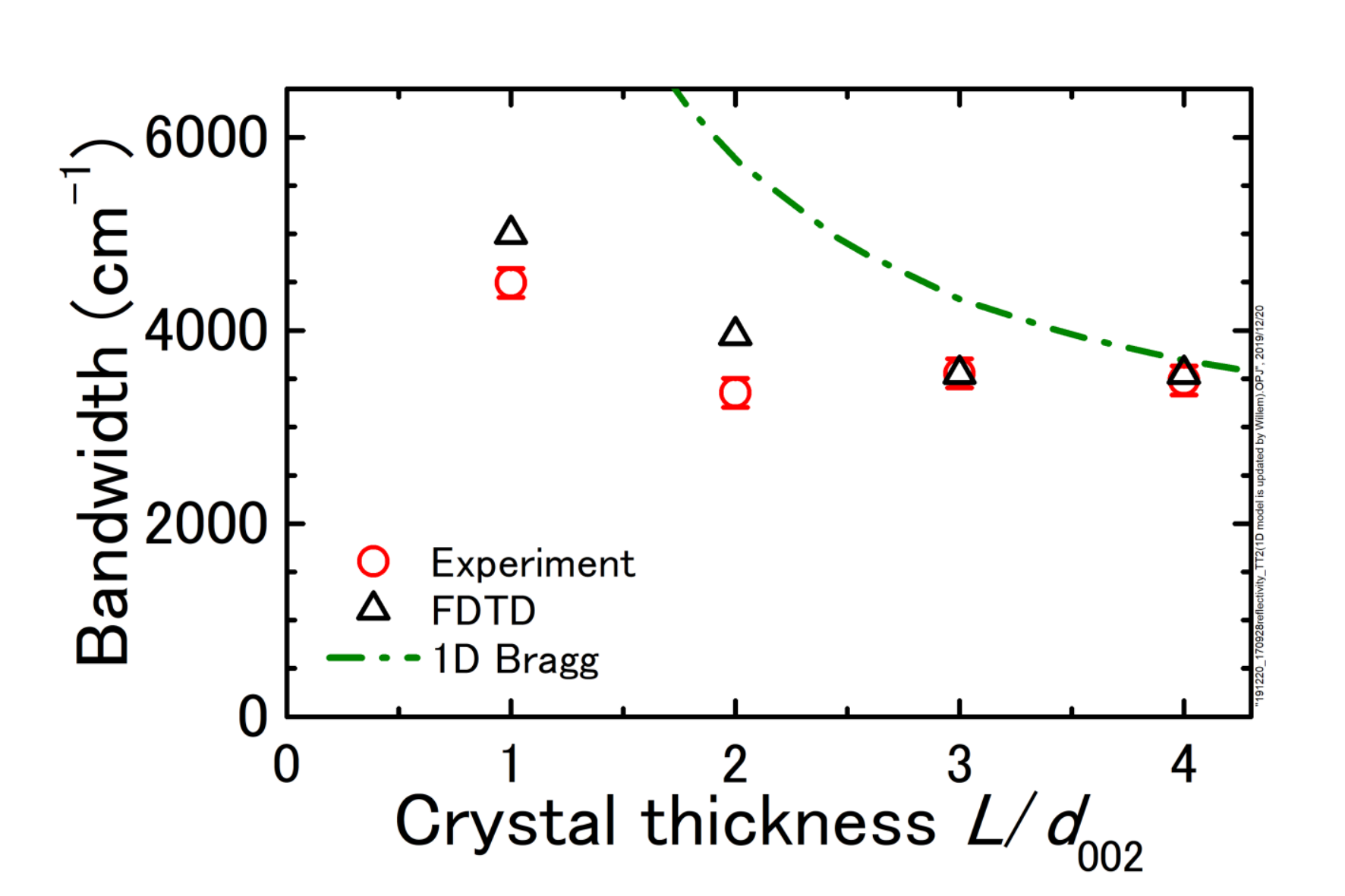}
\caption{
%%{\color{blue}WV: $L \rightarrow \infty$ bandwidth is not correct; would bandstructure from FDTD solve this? Are structure parameter in both FDTD and PW really the same?} 
Bandwidth of the stop bands (full width at half maximum of the reflectivity peak) versus crystal thickness reduced by the lattice spacing $(L/d_{002})$. 
The red circles with error bars are the experimental observations, the black triangles are calculated by 3D FDTD simulations. 
The green curve is the model without free parameters given in Eqs.~(\ref{eq:1D_Bragg},~\ref{eq:finite_width}) for weakly interacting photonic crystals with only one reciprocal lattice vector $\mathbf{g}$ as shown in Fig.~\ref{fig:cartoon-Bragg}(a). 
} 
\label{fig:bandwidth}
\end{figure}
%%%%%%%%%%%%%%%%%%%%%%%%%%

%%%%%%%%%%%%%%%%%%%%%%%%%%
\section{Discussion}\label{sec:discussion}
%%%%%%%%%%%%%%%%%%%%%%%%%%
%%%%%%%%%%%%%%%%%%%%%%%%%%%%%%%%%%
\subsection{Multiple Bragg diffraction}
%%%%%%%%%%%%%%%%%%%%%%%%%%%%%%%%%%
In order to investigate finite size effects in our 3D crystals, we estimated the stop band widths from the reflectivity spectra. 
Fig.~\ref{fig:bandwidth} shows the thickness dependence of the full width half maximum (FWHM) of the stop band for both the measured and the simulated reflectivity spectra. 
The stop band is slightly broader for the thinnest crystal thickness of $L = 2 d_{002}$ and the bandwidth only slightly decreases with increasing thickness. 
For a thickness of only $L = 3 d_{002}$ the bandwidth seems to have saturated to the infinite crystal limit. 
It is gratifying that the stop band widths obtained from the FDTD simulations agree very well with the experimentally obtained results, as shown in Fig.~\ref{fig:bandwidth}.

%%%%%%%%%%%%%%%%%%%%%%%%%%
\begin{figure}[h!]
\centering
\includegraphics[width=1.0\columnwidth]{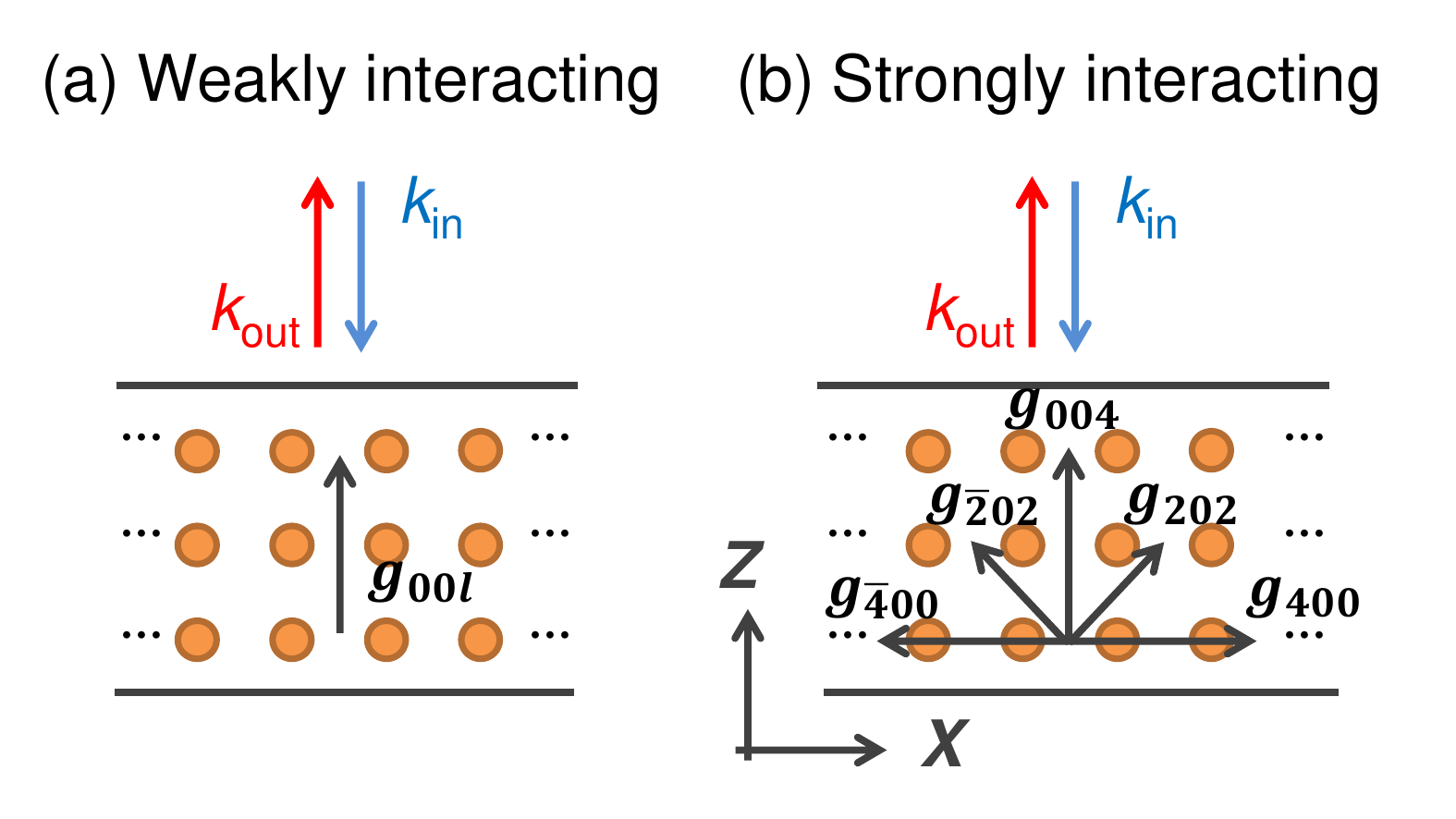}
\caption{
Schematic of Bragg conditions encountered in photonic crystal reflectivity experiments with incident wave vector $k_{in}$ (blue) and reflected wave vector $k_{out}$ (red).
The crystal is drawn as an $(x,z)$ cross-section. 
(a) In a weakly interacting photonic crystal, $k_{in}$ and $k_{out}$ only excite Bragg waves with a reciprocal lattice vector ($\mathbf{g}_{00l}$) parallel to the incident and reflected waves. 
(b) In a strongly interacting photonic crystal, $k_{in}$ and $k_{out}$ excite multiple Bragg diffraction conditions simultaneously with reciprocal lattice vectors that are not only parallel to the incident and reflected waves (here $\mathbf{g}_{00l} = \mathbf{g}_{004}$), but also reciprocal lattice vectors in oblique directions, clockwise from left: $\mathbf{g}_{\bar400}$, $\mathbf{g}_{\bar202}$, $\mathbf{g}_{202}$, $\mathbf{g}_{400}$. 
}
\label{fig:cartoon-Bragg}
\end{figure}
%%%%%%%%%%%%%%%%%%%%%%%%%%

To obtain a physical intuition of the stop band widths of the 3D photonic band gap crystals, we start from the basic notion that light propagation in photonic crystals is in essence determined by Bragg diffraction interference~\cite{Joannopoulos2008Book}. 
The thickness dependence of the stop band widths for the 3D photonic crystals in Fig.~\ref{fig:bandwidth} stems from intricacies in the Bragg conditions. Figure~\ref{fig:cartoon-Bragg} (a) illustrates the well-known Bragg diffraction~\cite{Ashcroft1976Book} where incident light with a wave vector $k_{in}$ is diffracted to outgoing light waves with wave vector $k_{out}$ by the crystal periodicity that is described by a single dominant reciprocal lattice vector $\mathbf{g}_{00l}$, such that 
%%%%%%%%%%%%%%%%
\begin{equation}
k_{out} - k_{in} = \mathbf{g}_{00l}.
\label{eq:simple_Bragg_condition}
\end{equation}
%%%%%%%%%%%%%%%%
This simple Bragg condition leads to the formation of a stop gap for light incident in the $k_{in}$ direction~\cite{Yariv1980book}. 
It appears that for photonic crystals with a low photonic strength such as opals and colloidal crystals~\cite{Bertone1999PRL, Hartsuiker2008thesis}, the observed total relative stopband width $\Delta \omega/\omega_c$ - where $\omega_c$ is the center frequency of the stop band - is well described by the following heuristic expression~\cite{Bechger2003thesis, Hartsuiker2008thesis}
%%%%%%%%%%%%%%%%%
\begin{equation}\label{eq:1D_Bragg}
\frac{\Delta \omega}{\omega_c} = 
\sqrt{\left(\frac{\Delta \omega}{\omega_c}\right)^{2}_{phot} + 
\left(\frac{\Delta \omega}{\omega_c}\right)^{2}_{fin}}, 
\end{equation}
%%%%%%%%%%%%%%%%%
where $\left(\Delta \omega/\omega_c\right)_{phot}$ is the purely photonic bandwidth that is equal to the relative width of the dominant stop gap in a band structure. 
Since a band structure pertains to an infinitely extended crystal ($L \rightarrow \infty$)~\footnote{The relative bandwidth of the stop band in the dominant crystal direction is a practical measure for the photonic interaction strength between the light and the photonic crystal~\cite{Vos1996JPCM,Vos2015book}.}, the relative bandwidth $\left(\Delta \omega/\omega_c\right)_{phot}$ is constant. 
When the photonic width in Eq.~\ref{eq:1D_Bragg} is taken to only depend on one reciprocal lattice vector ($n = 1$), we refer to it as one-dimensional (1D) model. 
Secondly, $\left(\Delta \omega/\omega_c\right)_{fin}$ is the relative bandwidth caused by the finite thickness of the crystal; a practical expression for this thickness dependence is well-known in X-ray diffraction theory and in grating theory to be proportional to a sinc function that is proportional to the inverse thickness, see Ref.~\cite{Warren1969book}, Eq.~(3.6): 
%%%%%%%%%%%%%%%%
\begin{equation}
\left(\frac{\Delta \omega}{\omega_c}\right)_{fin} = C~\frac{d_{002}}{L}, 
\label{eq:finite_width} 
\end{equation}
%%%%%%%%%%%%%%%%
where in case of the FWHM of a sinc function the constant is to a very good approximation equal to $C = 1.4$. 
Figure~\ref{fig:bandwidth} shows that the 1D model consisting of Eqs.~\ref{eq:1D_Bragg} and \ref{eq:finite_width} has a much steeper dependence than both our measured data and our simulated FDTD results. 
Therefore, we conclude that the 1D Bragg model is not a suitable description of our observations. 
%endblue

Here, we focus on 3D photonic crystals with a band gap, therefore with a high photonic strength. 
In case of a high photonic strength, as illustrated in Fig.~\ref{fig:cartoon-Bragg}(b), Bragg interference becomes more complex compared to the situation in Fig.~\ref{fig:cartoon-Bragg}(a). 
It appears that the first, simple, Bragg diffraction (similar to the one in Fig.~\ref{fig:cartoon-Bragg}(a)) is equivalent to multiple Bragg conditions~\cite{vanDriel2000PRB} 
%%, where sums of reciprocal lattice vectors appear that are equal the one in the simple Bragg diffraction
%%%%%%%%%%%%%%%%%%%%%%%%%%%%%%%%%%%%
\begin{align}
k_{out} - k_{in} &=  \mathbf{g}_{00l} = \mathbf{g}_{004} \label{eq:multiple_Bragg2}\\
        &= \mathbf{g}_{\bar202} + \mathbf{g}_{202} \label{eq:multiple_Bragg3}\\ 
        &= \mathbf{g}_{400} + \mathbf{g}_{\bar202} + \mathbf{g}_{\bar202} \label{eq:multiple_Bragg4}\\
        &= \mathbf{g}_{\bar400} + \mathbf{g}_{202} + \mathbf{g}_{202} \label{eq:multiple_Bragg5}\\
        &= \mathbf{g}_{040} + \mathbf{g}_{0\bar22} + \mathbf{g}_{0\bar22} \label{eq:multiple_Bragg6}\\
        &= \mathbf{g}_{0\bar40} + \mathbf{g}_{022} + \mathbf{g}_{022} \label{eq:multiple_Bragg7}
\end{align}
%%%%%%%%%%%%%%%%%%%%%%%%%%%%%%%%%%%%
Since our 3D photonic crystals have the diamond structure, the first, simple, Bragg diffraction in the $z$-direction is the one with Miller indices $hkl = 00l = 004$. 
The vectorial sum of the reciprocal lattice vectors in Eqs.~(\ref{eq:multiple_Bragg3},~\ref{eq:multiple_Bragg4},~\ref{eq:multiple_Bragg5},~\ref{eq:multiple_Bragg6},~\ref{eq:multiple_Bragg7}) add up to the one in Eq.~(\ref{eq:multiple_Bragg2}), as is appreciated from the sum of their Miller indices adding up to those of the first one. 
Such equivalent reciprocal lattice vectors in the diamond structure are illustrated in Fig.~\ref{fig:cartoon-Bragg}(b). 
The consequence from Eqs.~(\ref{eq:multiple_Bragg3},~\ref{eq:multiple_Bragg4},~\ref{eq:multiple_Bragg5},~\ref{eq:multiple_Bragg6},~\ref{eq:multiple_Bragg7}) is that the reciprocal vectors in the multiple Bragg processes are not oriented in the direction of the thinnest crystal dimension ($z$-direction) but are also oriented in oblique directions, in the extreme case even in the perpendicular directions $x$ and $y$. 
In these directions, the crystal is much more extended. 
Therefore, these Bragg conditions are hardly affected by the finite-size effects in the $z$-direction. 
Since the large number of oblique Bragg conditions also contribute to the overall bandwidth of the stop band that is observed in the $k_{out}$ direction, it is thus sensible that the observed bandwidth depends only little on the finite thickness in the $z$-direction, in agreement with the observed experimental results in Fig.~\ref{fig:bandwidth}. 

In parallel to the bandwidth, the maximum reflectivity of the stop band also hardly varies with increasing crystal thickness. 
For the experiment, this is apparent from the  observations in Fig.~\ref{fig:reflectivity_all_spectra}. 
This result can also be rationalized with the insight that interference of many oblique Bragg conditions contribute to the observed reflectivity peak.
Since these oblique Bragg conditions experience only little finite size effects, on account of the lateral extent of the crystal being much larger than the Bragg length, they contribute maximal amplitudes to the whole reflectivity peak regardless of the 3D crystal thickness. %}%%endblue 
Therefore, we conclude that the diffraction of strongly interacting photonic crystals - certainly those with a complete 3D photonic band gap - is not only determined by the reciprocal lattice vectors in the direction of the incident and diffracted wave vectors, but by \textit{all} reciprocal lattice vectors.

%%%%%%%%%%%%%%%%%%%%%%%%%%%%%%%%%%
{\subsection{Practical consequences}
%%%%%%%%%%%%%%%%%%%%%%%%%%%%%%%%%%
Our work aims to contribute to a deeper understanding of real 3D photonic crystals with finite size. 
The conclusion so far is that thin 3D photonic band gap crystals with high photonic strength already yield potent transport behavior thanks to multiple Bragg interference, which is highly useful from an application point of view.
The underlying reason why thin photonic band gap crystals are potent is the bandwidth of the band gap: broad band gaps are sustained by thin structures (whereas narrow band gaps, as with inverse opals, require thick structures.) 
Such broad 3D photonic band gaps are extremely useful for applications for a number of reasons: 
Firstly, broad photonic band gaps serve to process broad signal bandwidths, which is useful for optical communication purposes.
Secondly, in quantum information applications, elementary quantum systems such as two-level systems are more strongly shielded from external vacuum noise~\cite{Clerk2010RMP} when a surrounding 3D photonic band gap is broader.
Thirdly, a broader band gap is more robust to disorder~\cite{Chutinan1999JOSAB, Li2000PRB, Koenderink2005PRB, Woldering2009JAP}, thus making a band gap device more robust to inadvertent fabrication errors. 
Fourthly, a smaller photonic band gap device is still functional with a broader 3D band gap, since the Bragg length is then even shorter, thus permitting smaller sizes. 
Thus, we believe that our present study on finite size effects for broad band gaps is relevant for various applications using 3D photonic crystals. 
Future work could address the local density of state (LDOS) and the density of states (DOS)~\cite{Barnes2019arxiv} in a 3D photonic band gap, as they play central roles in spontaneous emission control and the control of blackbody processes.
}

%%%%%%%%%%%%%%%%%%%%%%%%%%%%%
\section{Summary and outlook}
%%%%%%%%%%%%%%%%%%%%%%%%%%%%%
In this study, we have investigated the optical reflectivity of three-dimensional (3D) photonic band gap crystals as increasing thickness. 
The 3D crystals with the so-called woodpile structure are assembled by stacking GaAs plates with nanorod arrays using an advanced micro-manipulation method. 
We observe broad and intense stop bands originating in a stop gap along the stacking direction, typical of strongly interacting photonic crystals. 
The bandwidth and the maximum reflectivity of the stop band hardly increase with increasing crystal thickness. 
This weak thickness dependence differs remarkably from previous reports on weakly interacting 3D photonic crystals where stop band widths and peaks show a strong thickness dependence. 
Our current results are well matched by FDTD simulations. 
Therefore, we conjecture that multiple Bragg interference relevant to several reciprocal lattice vectors including oblique ones contribute to the formation of the stop band in our 3D crystals, which demonstrates a finite size effect in the strongly interacting photonic crystals.  
This work contributes to a deeper understanding of strongly interacting 3D photonic crystals with finite size, and paves the way to the applications of 3D photonic band gap crystals.

%%%%%%%%%%%%%%%%%%%%%%%%%%%%%
\section{Acknowledgments}
%%%%%%%%%%%%%%%%%%%%%%%%%%%%%
We thank Diana Grishina, Pepijn Pinkse, Ad Lagendijk for useful discussions. 
We acknowledge funding by KAKENHI (15H05700, 15H05868, 17H02796), FOM-NWO program "Stirring of light!", STW-Perspectief program "Free-form scattering optics", and the MESA+ section Applied Nanophotonics (ANP). 

%%%%%%%%%%%%%%%%%%%%%%%%%%%%%
\section*{Appendix A}\label{sec:appendix}
%%%%%%%%%%%%%%%%%%%%%%%%%%%%%
%%{\color{red}
Our 3D photonic crystal (shown in Fig.~\ref{fig:schematic_structure} (a)) does not have mirror symmetry with respect to any x-y plane. 
Thus, in addition to the thickness, the order of stacked layers in the 3D crystals also affects reflectivity spectra, although the effect is relatively minor. 
Here, we note the stacking order of all 3D crystal samples with the thicknesses $L = d_{002}$, $2d_{002}$, $3d_{002}$ and $4d_{002}$, shown in Fig.~\ref{fig:SEM_of_the_crystals}(a). 

As explained in Sect. II A., the design of the 3D crystals are composed of the periodically stacked four layers with arrays of nanorods. 
In the two of these four layers, the nanorods are parallel to $x$ axis, and in the other two, parallel to $y$ axis. 
Also, the nanorods are spatially shifted by $d/2$ between the two layers with parallel nanorods. 
We label these four layers as Layer 1, 2, 3 and 4.
The rod patterns of these layers are defined below:\\
Layer 1: without shift and parallel to the y-axis. \\
Layer 2: without shift and parallel to the x-axis. \\
Layer 3: with shift and parallel to the y-axis. \\
Layer 4: with shift and parallel to the x-axis. 

We fabricated these rod patterns in square regions of plates as shown in Fig.~\ref{fig:SEM_of_the_crystals} (a). 
In the square region, the pattern without the shift correspond to having a rod at the center of the square region, and that with the shift a rod at the position shifted by $d/2$ from the center along the direction perpendicular to the nanorods.
These layers are stacked, following the order below for making samples with different thickness $L$:\\ 
$L = 1 d_{002}$: 1-4\\
$L = 2 d_{002}$: 1-4-3-2 \\
$L = 3 d_{002}$: 1-4-3-2-1-4 \\
$L = 4 d_{002}$: 2-3-4-1-2-3-4-1 \\
These numbers represent the layer number, and shown in order from the left to the right for layer numbers from the top to the bottom of the structure.
The order of stacking was inadvertently modified for the thickness $L = 4 d_{002}$, resulting in the nanorods of the top layer perpendicular to the electric field of the incident light.
%In particular, the relationship between the length direction of nanorods and the polarization direction of incident light results in a minor variation in reflectivity spectra in Fig.~\ref{fig:reflectivity_spectra}. 
%%}%%end-red

%%TO bE FINISHED...: 
%%DEFINITION of layers in the structure: \\
%%Layer 1, nanorods parallel to the y-axis $\&$ at the center. \\
%%Layer 2, nanorods parallel to the x-axis $\&$ at the center. \\
%%Layer 3, nanorods parallel to the y-axis $\&$ NOT at the center. \\
%%Layer 4, nanorods parallel to the x-axis $\&$ NOT at the center. \\
%%Structure 2 layers: 1-4 (top to bottom)\\
%%Structure 4 layers: 1-4-3-2 \\
%%Structure 6 layers: 1-4-3-2-1-4 \\
%%Structure 8 layers: 2-3-4-1-2-3-4-1 \\
%%E-field is parallel to the long axis of the trenches (Fig. 3(a)) which is parallel to x. 
%%WLV: move to appendix or methods

%%%%%%%%%%%%%%%%%%%%%%%%%%%%%%%%%%%%%%%%%%%%%%%%%%%%%%%%

%%%%%%%%%%%%%%%%%%%%%%%%%%%%%%%%%%%%%%%%%%%%%%%%%%%%%%

%%%%%%%%%%%%%%%%%%%%%%%%%%%%%%%%%%%%%%%%%%%%%%%%%%%%%%%

\begin{thebibliography}{} 
% Please use as label: NameYearJournal, e.g., Tajiri2015APL. Thanks! 
% NameYearJournal
\bibitem{Bykov1972SovPhysJETP}
V. P. Bykov, 
``Spontaneous emission in a periodic structure", 
Soviet Phys. JETP \textbf{35}, 269-- (1972).

\bibitem{Yablonovitch1987PRL} 
E. Yablonovitch,
``Inhibited Spontaneous Emission in Solid-State Physics and Electronics,"
Phys. Rev. Lett. \textbf{58}, 2059--2062 (1987).

\bibitem{John1987PRL}
S. John, 
``Strong Localization of Photons in Certain Disordered dielectric superlattices,"
Phys. Rev. Lett. \textbf{58}, 2486--2489 (1987).

\bibitem{John1990PRL}
S. John and J. Wang, 
``Quantum Electrodynamics near a Photonic Band Gap: Photon Bound States and Dressed Atoms,"
Phys. Rev. Lett. \textbf{58}, 2418--2421 (1990).

%%\bibitem{Joannopoulos1997Nature}
%%J. D. Joannopoulos, P. R. Villeneuve, and S. Fan, 
%%``Photonic crystals: putting a new twist on light,"
%%Nature (London) \textbf{386}, 143 (1997).

\bibitem{Lopez2003AdvMater}
C. L{\'o}pez,
``Materials Aspect of Photonic Crystals",
Adv. Mater. \textbf{15}, 1679--1704 (2003).

\bibitem{Lourtioz2008Book}
J.-M. Lourtioz, H. Benisty, V. Berger, J.-M. G{\'e}rard, D. Maystre, and A. Tchelnokov, 
\textit{Photonic Crystals: Towards Nanoscale Photonic Devices}
(Springer, New York, 2008) 2nd Ed. 

\bibitem{Joannopoulos2008Book}
J.~D. Joannopoulos, S.~G. Johnson, J.~N. Winn, and R.~D. Meade, 
\textit{Photonic Crystals: Molding the Flow of Light} 
(Princeton University Press, Princeton NJ, 2008) 2nd Ed.

\bibitem{Ghulinyan2015Book}
M. Ghulinyan and L. Pavesi, Eds., 
\textit{Light Localisation and Lasing: Random and Pseudorandom Photonic Structures} 
(Cambridge Univ. Press, Cambridge, 2015).

\bibitem{Ashcroft1976Book}
N.~W. Ashcroft and N.~D. Mermin,
\textit{Solid State Physics}  
(Holt, Rinehart, and Winston, New York, 1976).

\bibitem{Lin1998Nature}
S.~Y. Lin, J.~G. Fleming, D.~L. Hetherington, B.~K. Smith, R. Biswas, K.~M. Ho, M.~M. Sigalas, W. Zubrzycki, S.~R. Kurtz, and J. Bur, 
``A three-dimensional photonic crystal operating at infrared wavelengths,” 
Nature \textbf{394}, 251–-253 (1998).

\bibitem{Thijssen1999PRL}
M.~S. Thijssen, R. Sprik, J.~E.~G.~J. Wijnhoven, M. Megens, T. Narayanan, A. Lagendijk, and W.~L. Vos, 
``Inhibited light propagation and broadband reflection in photonic air-sphere crystals,” 
Phys. Rev. Lett. \textbf{83}, 2730-–2733 (1999).

\bibitem{Noda1999APL} 
S. Noda, N. Yamamoto, H. Kobayashi, M. Okano, and K. Tomoda,
``Optical properties of three-dimensional photonic crystals based on III–V semiconductors at infrared to near-infrared wavelengths," 
Appl. Phys. Lett. \textbf{75}, 905--907 (1999). 

\bibitem{Noda2000Science} 
S. Noda, K. Tomoda, N. Yamamoto, and A. Chutinan, 
``Full Three-Dimensional Photonic Bandgap Crystals at Near-Infrared Wavelengths," 
Science \textbf{289}, 604--606 (2000). 

\bibitem{Blanco2000Nature}
A. Blanco, E. Chomski, S. Grabtchak, M. Ibisate, S. John, S. W. Leonard, C. L{\'o}pez, F. Meseguer, H. M{\'i}guez, J.~P. Mondia, G. A. Ozin, O. Toader, and H.~M. van Driel,
``Large-scale synthesis of a silicon photonic crystal with a complete three-dimensional bandgap near 1.5 micrometres,”
Nature \textbf{405}, 437--440 (2000).

\bibitem{Vlasov2000Nature}
Y.~A. Vlasov, X.-Z. Bo, J. C. Sturm, and D. J. Norris,
``On-chip natural assembly of silicon photonic bandgap crystals,”
Nature \textbf{414}, 289-–293 (2001).

\bibitem{Palacios-Lidon2002APL}
E. Palacios-Lid{\'o}n, A. Blanco, M. Ibisate, F. Meseguer, C. L{\'o}pez, and J.~J. S{\'a}nchez-Dehesa, 
``Optical study of the full photonic band gap in silicon inverse opals,” 
Appl. Phys. Lett. \textbf{81}, 4925-–4927 (2002).

\bibitem{Aoki2003NatMater} 
K. Aoki, H.~T. Miyazaki, H. Hirayama, K. Inoshita, T. Baba, K. Sakoda, N. Shinya, and Y. Aoyagi, 
``Microassembly of semiconductor three-dimensional photonic crystals," 
Nat. Mater. \textbf{2}, 117--121 (2003). 

\bibitem{Schilling2005APL}
J. Schilling, J. White, A. Scherer, G. Stupian, R. Hillebrand, and U. G{\"o}sele, 
``Three-dimensional macroporous silicon photonic crystal with large photonic band gap,” 
Appl. Phys. Lett. \textbf{86}, 011101 (2005).

\bibitem{Garcia2007AM}
F. García-Santamar{\'i}a, M. Xu, V. Lousse, S. Fan, P. V. Braun, and J. A. Lewis, 
``A germanium inverse woodpile structure with a large photonic band gap,” 
Adv. Mater. \textbf{19}, 1567–1570 (2007).

\bibitem{Subramania2007OE}
G. Subramania, Y. J. Lee, I. Brener, T. Luk, and P. Clem,
``Nano-lithographically fabricated titanium dioxide based visible frequency three dimensional gap photonic crystal,”
Opt. Express \textbf{15}, 13049–13057 (2007).

\bibitem{Takahashi2009NatMat}
S. Takahashi, K. Suzuki, M. Okano, M. Imada, T. Nakamori, Y. Ota, K. Ishizaki, and S. Noda, 
``Direct creation of three-dimensional photonic crystals by a top-down approach,” 
Nat. Mater. \textbf{8}, 721–725 (2009).

\bibitem{Staude2010OL}
I. Staude, M. Thiel, S. Essig, C. Wolff, K. Busch, G. von Freymann, and M. Wegener, 
``Fabrication and characterization of silicon woodpile photonic crystals with a complete bandgap at telecom wavelengths,” 
Opt. Lett. \textbf{35}, 1094--1096 (2010). 

\bibitem{Huisman2011PRB} 
S.~R. Huisman, R.~V. Nair, L.~A. Woldering, M.~D. Leistikow, A.~P. Mosk, and W.~L. Vos, 
``Signature of a three-dimensional photonic band gap observed on silicon inverse woodpile photonic crystals," 
Phys. Rev. B \textbf{83}, 205313: 1-7 (2011). 

\bibitem{Subramania2011NL}
G. Subramania, Q. Li, Y. J. Lee, J. J. Figiel, G. T. Wang, and A. J. Fische, “Gallium nitride based logpile photonic crystals,” Nano Lett. \textbf{11}, 4591–4596 (2011).

\bibitem{Frolich2013AM}
A. Fr{\"o}lich, J. Fischer, T. Zebrowski, K. Busch, and M. Wegener, 
``Titania woodpiles with complete three-dimensional photonic bandgaps in the visible,” 
Adv. Mater. \textbf{25}, 3588--3592 (2013).

\bibitem{Marichy2016SR}
C. Marichy, N. Muller, L. S. Froufe-P{\'e}rez, and F. Scheffold, 
``High-quality photonic crystals with a nearly complete band gap obtained by direct inversion of woodpile templates with titanium dioxide,” 
Sci. Rep. \textbf{6}, 21818 (2016).

\bibitem{Adhikary2019Arxiv}
M. Adhikary, R. Uppu, C.~A.~M. Harteveld, D.~A. Grishina, and W.~L. Vos, 
``Experimental probe of a complete 3D photonic band gap," 
ArXiv.org/abs/1909.01899 (2019). 

\bibitem{Yariv1980book}
A. Yariv and P. Yeh, 
\textit{Optical Waves in Crystals: Propagation and Control of Laser Radiation} 
(Wiley, New York, 1980) chapter 6, p. 155--. 

\bibitem{Vos2015book}
W.~L. Vos and L.~A. Woldering, 
in: \textit{Light Localisation and Lasing}, 
Eds. M. Ghulinyan and L. Pavesi 
(Cambridge University Press, Cambridge, 2015) Ch. 8, pp. 180--196, available as {ArXiv.org/abs/1504.06803}. 

\bibitem{Vos1996JPCM}
W.~L. Vos, M. Megens, C.~M. van Kats, and P. B\"osecke, 
``Transmission and diffraction by photonic colloidal crystals," 
J. Phys.: Condens. Matter \textbf{8}, 9503--950 (1996).

\bibitem{Lambropoulos2000RP}
P. Lambropoulos, G.~M. Nikolopoulos, T.~R. Nielsen, and S. Bay, 
``Fundamental quantum optics in structured reservoirs," 
Rep. Prog. Phys. \textbf{63}, 455-–503 (2000).

\bibitem{Smith1998MOTL} 
G. S. Smith, M. P. Kesler, and J. G. Maloney, 
``Dipole antennas used with all-dielectric, woodpile photonic-bandgap reflectors: Gain, field patterns, and input impedence,” 
Microw. Opt. Technol. Lett. \textbf{21}, 191--196 (1998).

\bibitem{Bermel2007OE}
P. Bermel, C. Luo, L. Zeng, L. C. Kimerling, and J.~D. Joannopoulos, 
``Improving thin-film crystalline silicon solar cell efficiencies with photonic crystals,” 
Opt. Express \textbf{15}, 16986--17000 (2007).

\bibitem{Wehrspohn2012JO}
R. B. Wehrspohn and J. \"Upping, 
``3D photonic crystals for photon management in solar cells,” J. Opt. \textbf{14}, 024003 (2012).

\bibitem{Koenderink2015science}
A. F. Koenderink, A. Alú, and A. Polman, 
``Nanophotonics: Shrinking light-based technology,” 
Science \textbf{348}, 516--521 (2015).

\bibitem{Nelson2011NM}
E. C. Nelson, N. L. Dias, K. P. Bassett, S. N. Dunham, V. Verma, M. Miyake, P. Wiltzius, J. A. Rogers, J. J. Coleman, X. Li, and P. V. Braun, 
``Epitaxial growth of three-dimensionally architectured optoelectronic devices,” 
Nat. Mater. \textbf{10}, 676--681 (2011).

\bibitem{David2012RPP}
A. David, H. Benisty, and C. Weisbuch, 
``Photonic crystal light-emitting sources,” 
Rep. Prog. Phys. \textbf{75}, 126501 (2012).

\bibitem{Li2003JOSA} 
Z.-Y. Li and K.-M. Ho, 
``Waveguides in three-dimensional layer-by-layer photonic crystals,” 
J. Opt. Soc. Am. B \textbf{20}, 801-–809 (2003). 

\bibitem{Staude2011OptLett}
I. Staude, G. von Freymann, S. Essig, K. Busch and M. Wegener, 
``Waveguides in three-dimensional photonic-bandgap materials by direct laser writing and silicon double inverion," 
Opt. Lett. \textbf{36}, 67--69 (2011). 

\bibitem{Ishizaki2013NP}
K. Ishizaki, M. Koumura, K. Suzuki, K. Gondaira and S. Noda, 
``Realization of three-dimensional guiding of photons in photonic crystals," 
Nat. Photonics \textbf{7}, 133--137 (2013). 

\bibitem{Tajiri2019Optica} 
T. Tajiri, S. Takahashi, Y. Ota, K. Watanabe, S. Iwamoto, and Y. Arakawa, 
``Three-dimensional photonic crystal simultaneously integrating a nanocavity laser and waveguides," 
Optica \textbf{6}, 296--299 (2019).

\bibitem{Tandaechanurat2011NP} 
A. Tandaechanurat, S. Ishida, D. Guimard, M. Nomura, S. Iwamoto, and Y. Arakawa, 
``Lasing oscillation in a three-dimensional photonic crystal nanocavity with a complete bandgap,” 
Nat. Photonics \textbf{5}, 91--94 (2011).

\bibitem{Clerk2010RMP}
A. A. Clerk, M. H. Devoret, S. M. Girvin, F. Marquardt, and R. J. Schoelkopf, 
``Introduction to quantum noise, measurement, and amplification,” 
Rev. Mod. Phys. \textbf{82}, 1155--1208 (2010). 

\bibitem{Leistikow2011PRL}
M.~D. Leistikow, A.~P. Mosk, E. Yeganegi, S.~R. Huisman, A. Lagendijk, and W.~L. Vos, 
``Inhibited Spontaneous Emission of Quantum Dots Observed in a 3D Photonic Band Gap," 
Phys. Rev. Lett. \textbf{107}, 193903 (2011). 

\bibitem{wikipedia2018support}
See \url{https://en.wikipedia.org/wiki/Support_(mathematics)} (Oct. 16, 2018)

\bibitem{Hermann2002JOSAB}
C. Hermann and O. Hess, 
``Modified spontaneous-emission rate in an inverted-opal structure with complete photonic bandgap," 
J. Opt. Soc. Am. B \textbf{19}, 3013--3018 (2002).

\bibitem{Kole2003thesis}
J.~S. Kole, 
\textit{New methods for the numerical solution of Maxwell's equations} 
(Ph.D. thesis, University of Groningen, 2003).

\bibitem{Hasan2018PRL}
S.~B. Hasan, A.~P. Mosk, W.~L. Vos, and A. Lagendijk, 
``Finite-size Scaling of the Density of States in Photonic Band Gap Crystals," 
Phys. Rev. Lett. \textbf{120}, 237402 (2018). 

\bibitem{Bertone1999PRL}
J.~F. Bertone, P. Jiang, K.~S. Hwang, D.~M. Mittleman, and V.~L. Colvin, 
``Thickness Dependence of the Optical Properties of Ordered Silica-Air and Air-Polymer Photonic Crystals," 
Phys. Rev. Lett. \textbf{83}, 300--303 (1999). 

\bibitem{Hartsuiker2008Lang}
A. Hartsuiker and W.~L. Vos,
``Structural Properties of Opals Grown with Vertical Controlled Drying," 
Langmuir \textbf{24}, 4670--4675 (2008). 

\bibitem{Devashish2017PRB}
D. Devashish, S.B. Hasan, J.J.W. van der Vegt, and W.L. Vos,
``Reflectivity calculated for a three-dimensional silicon photonic band gap crystal with finite support,"
Phys. Rev. B, \textbf{95}, 155141: 1--17 (2017).

\bibitem{Ho1994SSC}
K.~M. Ho, C.~T. Chan, C.~M. Soukoulis, R. Biswas, and M. Sigalas,
``Photonic band gaps in three dimensions: New layer-by-layer periodic structures,"
Solid State Commun. \textbf{89}, 413--416 (1994)

\bibitem{Leung1993book}
K.~M. Leung, 
in "\textit{Photonic band gaps and localization}", NATO-ASI vol. 308, Ed. C. M. Soukoulis (Plenum Press, New York NY, 1993), pp. 269--281. 

\bibitem{Aoki2008NP} 
K. Aoki, D. Guimard, M. Nishioka, M. Nomura, S. Iwamoto, and Y. Arakawa, 
``Coupling of quantum dot light emission with a three-dimensional photonic crystal nanocavity," 
Nat. Photon. \textbf{2}, 688--692 (2008). 

\bibitem{Tajiri2015APL} 
T. Tajiri, S. Takahashi, Y. Ota, J. Tatebayashi, S. Iwamoto, and Y. Arakawa, 
``Demonstration of a three-dimensional photonic crystal nanocavity in a 110-layered diamond structure," 
Appl. Phys. Lett. \textbf{107}, 071102 (2015). 

\bibitem{Iwamoto2016Phot}
S. Iwamoto, S. Takahashi, T. Tajiri, and Y. Arakawa,
``Semiconductor Three-Dimensional Photonic Crystals with Novel Layer-by-Layer Structures,"  
Photonics \textbf{3}, 34 (2016).

\bibitem{Ctistis2010PRB}
G. Ctistis, A. Hartsuiker, E. van der Pol, J. Claudon, W.~L. Vos, and J.-M. G{\'e}rard, 
``Optical characterization and selective addressing of the resonant modes of a micropillar cavity with a white light beam," 
Phys. Rev. B \textbf{82}, 195330 (2010). 

\bibitem{Skauli2003JAP}
T. Skauli, P. S. Kuo, K. L. Vodopyanov, T. J. Pinguet, O. Levi, L. A. Eyres, J. S. Harris, M. M. Fejer, B. Gerard, L. Becouarn, and E. Lallier, 
``Improved dispersion relations for GaAs and applications to nonlinear optics," 
J. Appl. Phys. \textbf{94}, 6447 (2003).

%\bibitem{dispersion}
%See \url{http://luxpop.com/HU_v173.cgi?OpCode=73} (Aug. 3, 2019)

\bibitem{Koenderink2005PRB}
A. F. Koenderink, A. Lagendijk, and W. L. Vos, 
``Optical extinction due to intrinsic structural variations of photonic crystals," 
Phys. Rev. B \textbf{72}, 153102 (2005). 

\bibitem{vanDriel2000PRB} 
H. M. van Driel and W. L. Vos, 
``Multiple Bragg wave coupling in photonic band-gap crystals," 
Phys. Rev. B \textbf{62}, 9872--9875 (2000). 

\bibitem{Vos2000PLA}
W. L. Vos and H. M. van Driel, 
``Higher order Bragg diffraction by strongly photonic fcc crystals: onset of a photonic bandgap,"
Phys. Lett. A \textbf{272}, 101--106 (2000). 

\bibitem{Huisman2012PRL} 
S. R. Huisman, R. V. Nair, A. Hartsuiker, L. A. Woldering, A. P. Mosk, and W. L. Vos, 
``Observation of Sub-Bragg Diffraction of Waves in Crystals," 
Phys. Rev. Lett. \textbf{108}, 205313: 1--7 (2012).

\bibitem{Demtroder2006book}
W. Demtr{\"o}der, 
\textit{Laser Spectroscopy: Basic Concepts and Instrumentation} 
(Springer, Berlin, 2002, 3rd Edition). 

\bibitem{Euser2008PRB}
T. G. Euser, A. J. Molenaar, J. G. Fleming, B. Gralak, A.
Polman, and W. L. Vos, 
``All-optical octave-broad ultrafast switching of Si woodpile photonic band gap crystals,"
Phys. Rev. B, \textbf{77}, 115214 (2008).

\bibitem{Bechger2003thesis}
L. Bechger, 
\textit{Synthesis and fluorescence of opal and air-sphere photonic crystals} 
(PhD thesis, University of Twente, 2003, available from \url{www.photonicbandgaps.com})

\bibitem{Hartsuiker2008thesis} 
A. Hartsuiker, 
\textit{Ultrafast All-Optical Switching and Optical Properties of Microcavities and Photonic Crystals} 
(PhD thesis, University of Twente, 2008, available from \url{www.photonicbandgaps.com})

\bibitem{Chutinan1999JOSAB}
A. Chutinan and S. Noda, 
``Effects of structural fluctuations on the photonic bandgap during fabrication of a photonic crystal: a study of a photonic crystal with a finite number of periods,"
J. Opt. Soc. Am. B \textbf{16}, 1398--1402 (1999)

\bibitem{Li2000PRB}
Z.-Y. Li and Z.-Q. Zhang, 
``Fragility of photonic band gaps in inverse-opal photonic crystals," 
Phys. Rev. B. \textbf{62}, 1516--1519 (2000). 

\bibitem{Woldering2009JAP}
L.A. Woldering, A.P. Mosk,  R.W. Tjerkstra, and W.L. Vos, 
``The influence of fabrication deviations on the photonic band gap of three-dimensional inverse woodpile nanostructures,"
J. Appl. Phys. \textbf{105}, 093108 (2009).

\bibitem{Warren1969book}
B. E. Warren, 
\textit{X-ray Diffraction} 
(Addison-Wesley, Reading MA, 1969, reprinted by Dover, New York NY, 1990)

\bibitem{Vos1996PRB}
W.~L. Vos, R. Sprik, A. van Blaaderen, A. Imhof, A.
Lagendijk, and G.~H. Wegdam,
``Strong effects of photonic band structures on the diffraction of colloidal crystals," 
Phys. Rev. B. \textbf{53}, 16231--16235 (1996). 

\bibitem{Barnes2019arxiv}
W. L. Barnes, S. A. R. Horsley, and W. L. Vos, 
``Classical antennae, quantum emitters, and densities of optical states," 
arXiv.org/abs/1909.05619 (2019). 

\bibitem{ChangHasnain2012AOE}
C. J. Chang-Hasnain and W. Yang, 
``High-contrast gratings for integrated optoelectronics," 
Adv. Opt. Photon. \textbf{4}, 379–-440 (2012).

%%%%%%%%%%%%%%%%%%%%%%%%%%%%%%%%%%%%%%%%%%%%%%%%%%%%%%
\end{thebibliography}
\end{document}